\documentclass[epj]{webofc}
\usepackage[varg]{txfonts}   % Web of Conferences font
\usepackage{graphicx,color} % for figures
\usepackage{bm}       % bold math 
\usepackage{amsmath}  
\usepackage{amssymb}

\newcommand{\calO}{{\cal O}}

\newcommand\bfp{\mathbf p}

\newcommand\beq{\begin{eqnarray}}
\newcommand\eeq{\end{eqnarray}}

\newcommand\Fig[1]{Fig.~\ref{fig:#1}}
\newcommand\bal{\begin{align}}
\newcommand\eal{\end{align} }
\newcommand\Eq[1]{Eq.~\ref{eq:#1}}

\newcommand{\mybar}[1]%
        {\kern 0.6pt\overline{\kern -0.6pt#1\kern -0.6pt}\kern 0.6pt}
\woctitle{21st International Conference on Few-Body Problems in Physics}
\begin{document}
\title{Universal noise and Efimov physics}
\author{Amy N. Nicholson\inst{1}\fnsep\thanks{\email{anicholson@berkeley.edu}} }        % etc.

\institute{Department of Physics, University of California, Berkeley, CA, USA
          }

\abstract{
Probability distributions for correlation functions of particles interacting via random-valued fields are discussed as a novel tool for determining the spectrum of a theory. In particular, this method is used to determine the energies of universal $N$-body clusters tied to Efimov trimers, for even $N$, by investigating the distribution of a correlation function of two particles at unitarity. Using numerical evidence that this distribution is log-normal, an analytical prediction for the $N$-dependence of the $N$-body binding energies is made. 
}
\maketitle

Probability distributions are often studied in the context of Monte Carlo calculations, where a finite set of field configurations is used to estimate the value of an observable and its associated statistical uncertainty. Previously, it has been argued that the mean and variance of the distribution can often be estimated based on physical arguments, in efforts to better understand and control the signal-to-noise ratios in statistically noisy calculations \citep{Lepage:1989hd,Gibbs:1986xg}. Therefore, it seems plausible that this argument may be turned on its head, such that knowledge of the distribution can be used to make predictions about the theory. As an example, probability distributions of correlation functions in certain cases may prove to be useful tools for calculations of the spectrum of the theory. 

The concept of probability distributions arises naturally in lattice calculations. As an example, I will consider a Euclidean theory of non-relativistic particles interacting via a two-particle interaction, which will be mediated on the lattice by a Hubbard-Stratonovich field, $\phi$. 
The discretized action of the theory may be written $S=\psi^{\dagger} K \psi$, where, 
\beq K_{\bfp, \bfp'}(\tau,\tau') &=& \delta_{\tau,\tau'} D_{\bfp,\bfp'} + \delta_{\tau,\tau'-1} X_{\bfp,\bfp'}(\tau)\ , \cr D_{\bfp,\bfp'} &=& \delta_{\bfp,\bfp'} e^{\bfp^2/2M} \ , \qquad X_{\bfp,\bfp'}(\tau) = \delta_{\bfp,\bfp'} + \mathcal{C}^{1/2}(\bfp-\bfp') \tilde\phi_{\bfp-\bfp'}(\tau)\ ,
\label{eq:Dmat}
\eeq
\citep{Endres:2011er,Endres:2012cw}. Here $\tau$ denotes Euclidean time and $\tilde{\phi}_{\bfp}$ represents the Fourier transform of the $Z_2$-valued auxiliary field $\phi$. The limit of unitarity may be seen as a non-trivial UV fixed point in the $\beta$-function for the two-particle coupling, corresponding to a fine-tuning of the couplings in the lattice theory. In particular, one may define a tower of couplings which depend on powers of the momentum transfer, $C(\bfp) = \frac{4\pi}{M} \sum_{n=0}^{N_\calO-1} C_{2n} \calO_{2n}(\bfp)$, up to some cutoff $N_{\calO}$, and tune these to the unitary point, effectively removing the first $N_{\calO}$ orders of the effective range expansion for the scattering phase shift \citep{Endres:2011er}. 

In this formulation, the auxiliary fields are chosen to live along the temporal links of the lattice, so that, along with the choice of open temporal boundary conditions (valid only for zero temperature calculations), the matrix $K$ takes on an upper tridiagonal form. 
One important consequence of these choices is that the determinant, $\det K$ is simply the determinant of a product of free kinetic operators, $D$, which is independent of $\phi$, and therefore makes no contribution to the probability measure. Furthermore, propagators may be written as iterative operations in time, 
\beq
\label{eq:c1}
K^{-1}(\tau) = D^{-1} X(\tau-1) K^{-1}(\tau-1)
\eeq
with $K^{-1}(0) = D^{-1}$, which, for large Euclidean time, act as filters for the ground state,
\beq
\sum_{\phi} \langle \Psi | K^{-1}[\phi,\tau]| \Psi \rangle = \sum_n \langle \Psi | n \rangle e^{-E_n \tau} \langle n| \Psi \rangle \underset{\tau \to \infty}{\longrightarrow} \mathcal{Z} e^{-E_0 \tau} \ ,
\eeq
where $\sum_{\phi}$ denotes a sum over all field configurations $\phi$, and $E_n$ are single particle energies corresponding to the eigenstates $| n\rangle$ that have non-zero overlap onto the wavefunction $\Psi$ (the overlap factor onto the ground state is denoted by $\mathcal{Z}$). 

\begin{figure}
\centering
\sidecaption
\includegraphics[width=5cm,clip]{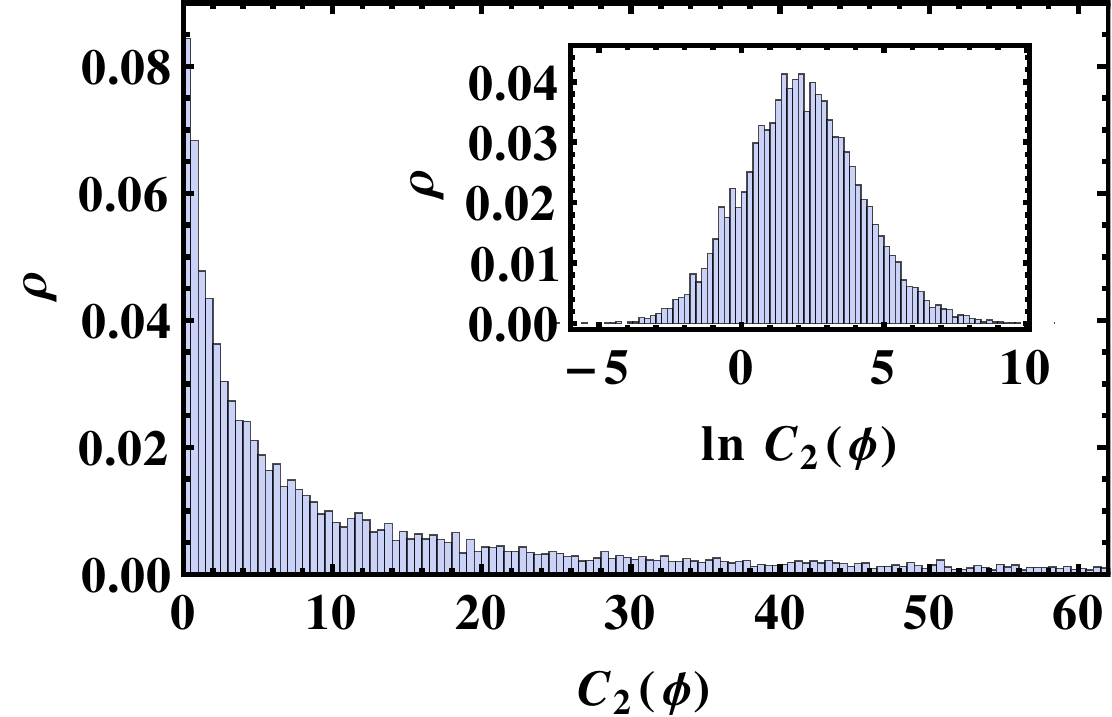}
\caption{\label{fig:hist}Histogram of the two-body correlator, $C_2$, at large Euclidean time, $\tau$. Inset: Histogram of the logarithm of the two-body correlator at the same $\tau$.}
\end{figure}

Given a finite set of field configurations, the probability distribution for an observable may be sampled by calculating the observable on each configuration (as is done in \Eq{c1} prior to performing the sum). The values of the observable on each configuration may be plotted as a histogram, with the shape of the histogram representing the distribution. For the two-particle correlator tuned to unitarity and calculated at large Euclidean time using the lattice theory described above, one finds the histogram shown in Fig.~\ref{fig:hist}. The inset shows the histogram obtained by calculating the logarithm of the observable, rather than the observable itself, on each sample. This distribution looks remarkably Gaussian, characteristic of an observable which obeys the log-normal distribution. 

While the lattice provides a natural tool for sampling probability distributions, the distributions themselves are not simply an artifact of lattice calculations, but may be defined field-theoretically in the continuum. Assuming we have a theory of particles interacting through a field, $\phi$, with probability measure $e^{-S[\phi]}$, then the probability distribution, $P$, and characteristic function, $\Phi$, for a functional $C_N[\phi]$, may be defined as
\beq
P(x) \propto \int [d\phi] e^{-S[\phi]}\delta (C_N[\phi]-x) \ , \quad \Phi(s) \propto \int [d\phi] e^{-S[\phi]+isC_N[\phi]} \ ,
\label{eq:char}
\eeq
The characteristic function is the Fourier transform of the probability distribution, and is itself a partition function for a theory that includes the observable in the Lagrangian, and may be amenable to perturbative treatments \citep{Grabowska:2012ik,Nicholson:2012xt}. In particular, $W(s) = -\ln \Phi(s)$ is the generating functional for the cumulants, $\kappa$, of the distribution,
\beq
W(s) = -\sum_{n=1}^{\infty} \frac{(is)^n}{n!} \kappa_n \ .
\eeq
Thus, if $C_N = \langle C_N[\phi] \rangle$ is an $N$-body correlation function (angle brackets denote an average over the field $\phi$ according to the appropriate probability measure for $\phi$) composed of $N$ propagators, then $\kappa_n$ may be identified as a sum of connected diagrams containing $n \times N$ external legs.

In this field theoretic language, the central limit theorem may be described in terms of renormalization group (RG) flow, the idea being that by successively block averaging random numbers drawn from a given distribution characterized by a set of cumulants, $\kappa_n$, one finds that the $n$th cumulant becomes rescaled as $\kappa_n \to N_b^{1-n/2} \kappa_n$ where $N_b$ is the number of block averaging operations (see, e.g., \citep{chaos}). Thus, cumulants with $n > 2$ act as irrelevant operators, vanishing in the limit $N_b \to \infty$.

The log-normal distribution follows the same RG arguments, however it applies to products of random numbers rather than sums. Thus, the log-normal distribution may be described as an RG fixed-point of products of random numbers. As described previously, propagators are given by $\tau$ products of $D^{-1}X[\phi]$, where the fields $\phi$ on each time slice are independent, random fields. Since we are interested in the limit $\tau \to \infty$, some RG argument along these lines may be applicable to this case. However, recall that $D^{-1}X[\phi]$ is a matrix, and little is known about products of random matrices above two-dimensions \citep{Jackson:2002qx}, so at this time a rigorous mathematical treatment along these lines is not possible.

One may understand the distribution, however, on physical grounds by examining the long-time behavior of the moments of the $N$-body correlation function. In the path integral language, the $n$th moment of the distribution, like the $n$th cumulant, is given by diagrams containing $n \times N$ external legs. The moments, however, include both connected and disconnected diagrams, and may be shown to correspond to $n \times N$-body correlation functions. As an example, if we begin with a two-body correlation function composed, for simplicity of this argument, of single particle propagators $\mathcal{S}[\phi,\tau]$,
\beq
C_2(\tau) = \langle \left( \mathcal{S}[\phi,\tau] \right)^2 \rangle \ ,
\eeq
then the variance of the probability distribution of this correlation function is given by the product of four propagators, averaged over a set of background fields,
\beq
\sigma_{C_2}^2(\tau) = \langle \left( \mathcal{S}[\phi,\tau] \right)^4 \rangle \ ,
\eeq
This quantity is, in turn, a four-body correlation function. Following this argument, one may easily verify that the $N$th moment corresponds to a $2N$-body correlation function. 

At this point it is important to note that in order for all moments to correspond to physical correlation functions, the formulation must be such that the particle content specified by the probability measure is not at odds with the particle content in the operator. For example, many theories include a fermion (or boson) determinant as part of the probability measure, with the power of the determinant determining the number of degenerate "flavors" of particle in the theory. As noted previously, determinants in this formulation do not contribute to the probability measure, and the particle content of the theory is fully determined by the operator (the set of propagators) for a given moment. In a non-relativistic theory this is in a sense natural because at zero temperature and zero chemical potential there are no particle/antiparticle pairs arising from the vacuum, so the only information about particle content must be encoded in the operators measured. 

Finally, because the long-time behavior of correlation functions is dominated by the ground-state energy of the system, the long-time behavior of the $N$th moment of the two-particle correlator, $\mathcal{M}_N(\tau)$. is then
\beq
\mathcal{M}_N(\tau) \underset{\tau \to \infty}{\longrightarrow} \mathcal{Z}_{2N} e^{-E_0^{(2N)} \tau} \ ,
\eeq
where $E_0^{(2N)}$ and $\mathcal{Z}_{2N}$ are the ground-state energy and wavefunction overlap of the $2N$-body system, respectively. It should be noted that the set of propagators in all cases are inherently symmetric under particle interchange, and will have overlap with both identical bosons and distinguishable particles. 

\begin{figure}
\centering
\sidecaption
\includegraphics[width=5cm,clip]{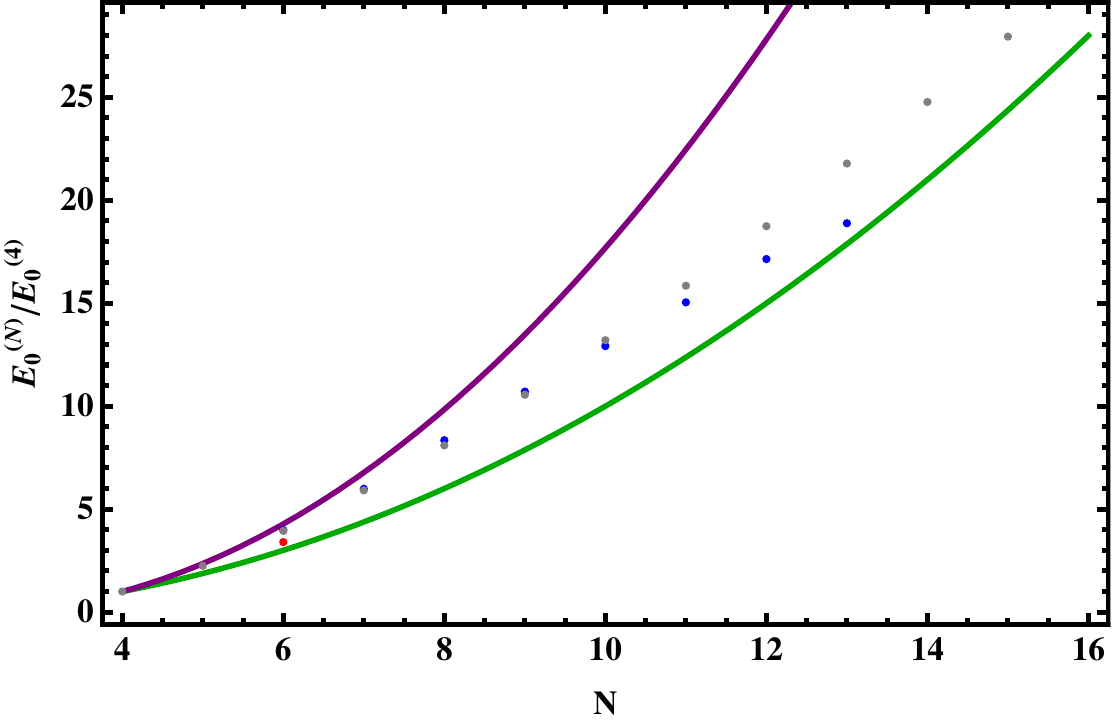}
\caption{\label{fig:comp}Ratio of energies $E_0^{(N)}/E_0^{(4)}$ predicted by the log-normal distribution (green), compared to direct numerical calculations of the energies, (blue) \citep{vonStecher:2009qw}, (red) \citep{2011PhRvL.107t0402V}, and (gray) \citep{2015arXiv150800081Y}, as well an ansatz (purple band) for the $N$ dependence of the energies based on direct numerical calculations \citep{Gattobigio:2013yda,Kievsky:2014dua}. The red point shows the only calculation involving an excited state, where non-universal effects are expected to be smaller than for ground-state calculations.}
\end{figure}

At unitarity, the energies of $N$-body bosonic systems have been shown to be tied to the spectrum of Efimov trimers \citep{V1970563,Efimov:1971zz}, whose binding energies are known to obey a discrete scaling symmetry, for at least $N\lesssim 16$ \citep{Blume201186,vonStecher:2009qw,Gattobigio:2013yda,Kievsky:2014dua,2015arXiv150800081Y}. One method for showing this is to study the ratio of some $N$-body binding energy to that of a trimer, while varying the short-distance behavior of the theory. If the ratio remains constant, this proves that there is no relevant $n$-body scale (for $3 < n \leq N$), meaning the three-body scale is the only scale determining the physics. This implies that for each trimer there must be a corresponding $N$-body state with the measured binding energy ratio, due to the discrete scale invariance. It also implies that this ratio is universal, in the sense that it is insensitive to any UV physics. These ratios have been calculated directly using numerical methods (see data in \Fig{comp}), however, at this time there is still notable disagreement among the various calculations.

It is now clear from the discussion that the energies of $N$-body bound states at unitarity should have the form $E_0^{(N)} = -a_{N} \Lambda_{*}$, where $\Lambda_{*}$ is a three-body scale that may be set using the energy of any of the Efimov trimers, and $a_N$ is a universal $N$-body energy ratio. Thus, the moments of the two-body correlation function may be shown to have the following form at large Euclidean time,
\beq
\mathcal{M}_N \underset{\tau \to \infty}{\longrightarrow} \mathcal{Z}_{2N} e^{a_{2N}\Lambda_{*} \tau} \ .
\label{eq:moments}
\eeq
The moments of the log-normal distribution are known to be,
\beq
\mathcal{M}_N^{(\mathrm{LN})} = e^{N \mu + \frac{1}{2}N^2 \sigma^2} \to \mathcal{M}_N^{(\mathrm{LN})} = e^{\frac{1}{2} N(N-1)\sigma^2} \ ,
\label{eq:lngenmoments}
\eeq
where $\mu$ and $\sigma$ are the mean and standard deviation of the distribution of the logarithm of the random variable, and on the right I have used the constraint that the first moment is a constant, corresponding to a correlation function with zero ground state energy. This form implies that the log-normal distribution corresponds to correlation functions whose moments are controlled by a single scale, since correlation functions decay as the exponential of an energy scale times Euclidean time. This is precisely what is expected from the form of the moments of the two-body correlation function at unitarity, \Eq{moments}. Furthermore, by comparing the log-normal moments with those of the two-body correlator at unitarity, one may predict the $N$-dependence of the universal constants, $a_{2N}$, giving the following energy ratios:
\beq
E_0^{(2N)}/E_0^{(4)} = \frac{1}{2} N(N-1) \ .
\eeq
One should note that the expression above is not expected to hold for arbitrarily large $N$, because as the states become increasingly bound, eventually they will become affected by non-universal cutoff effects. For the (zero effective range) lattice theory which is used to sample the distribution, the energy cutoff imposed by the lattice spacing is found to be a factor of $\sim 335$ times larger than the energy of the lowest Efimov trimer, allowing a large separation of scales in which to find universal behavior. 

The prediction based on the log-normal distribution is plotted in \Fig{comp} (lower solid line), along with a comparison to other works. The upper solid line is an analytic prediction from \citep{Gattobigio:2013yda,Kievsky:2014dua}, using data produced from a finite range potential, coupled with a procedure for removing the finite effective range effects. This conjecture confirmed the leading $N^2$ dependence predicted by this work, but with a significantly different prefactor. The points represent direct numerical calculations of the $N$-body binding energies using potential models with varying dependences on non-universal effects such as non-zero effective range and three- and higher-body cutoffs. Note that all of these calculations (including this work) are based on ground-state calculations, except for the single red point. For this point the binding energy of a 6-body state tied to the first excited Efimov trimer was calculated, and is therefore expected to be the least affected by non-universal UV effects. 

Finally, note that for odd numbers of particles the correlator is not positive so an exactly log-normal distribution would be somewhat unexpected. However, in practice it is found that the distributions for odd-numbered correlation functions are also approximately log-normal with only a tiny negative contribution. Because all direct numerical calculations of the $N$-body binding energies in other works show smooth behavior with $N$, I have interpolated the prediction for even $N$ in this plot.

To determine quantitatively how close the probability distribution for the two-body correlation function at unitarity is to a log-normal distribution, we may turn to the cumulant expansion \citep{Endres:2011jm}, which relates the logarithm of the (field averaged) correlation function to the sum of the cumulants of the distribution of $\ln C$,
\beq
\ln \langle C \rangle = \sum_{n=1}^{\infty} \frac{\kappa_n(\ln C)}{n!}  \ .
\label{eq:cumulantexp}
\eeq
Recall that the cumulants of a distribution act like connected $n$-point operators with increasing dimension. Thus the cumulant expansion may be regarded in an effective field theory context: as a distribution is driven toward log-normal, cumulants $\kappa_n(\ln C)$ with $n>2$ become increasingly irrelevant \citep{Endres:2011mm}. For the log-normal distribution, cumulants for $n>2$ are exactly zero; any deviation from log-normal may be quantified by the relative size of the third and higher cumulants.

\begin{figure}
\centering
\sidecaption
\includegraphics[width=5cm,clip]{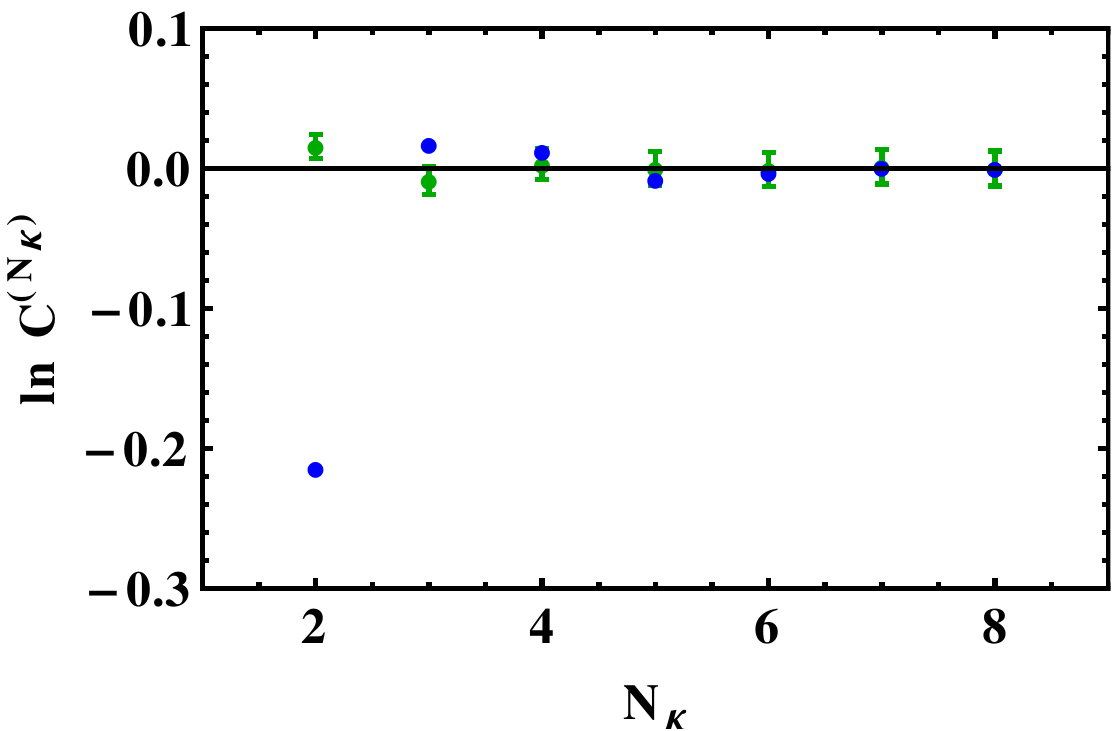}
\caption{\label{fig:exp}Cumulant expansion (\Eq{cumulantexp}) for the two-body correlator at unitarity cut off at $N_{\kappa}$. The green points represent the lattice data on a $16^3 \times 1000$ lattice, while the blue points are from a mock distribution created by expanding around unitarity in such a way as to reproduce moments corresponding to the energies found in \citep{vonStecher:2009qw}. }
\end{figure}

In \Fig{exp} the cumulant expansion for the logarithm of the two-body correlation function at large time is shown versus the number of cumulants included in the sum, $N_{\kappa}$. Comparing the value of the cumulant expansion cut off at $N_{\kappa} = 2$ to the asymptotic value, we see that the distribution deviates by approximately $2\%$ from the log-normal distribution. This small discrepancy is likely due to sensitivity to lattice artifacts for large moments of $C_2$. In the same figure an attempt has been made to quantitatively determine how much variations in the energies affect the cumulants of the distribution. An example of a mock distribution generated by expanding around log-normal and fitting the coefficients to produce the energies calculated in \citep{vonStecher:2009qw} is shown by the blue points. Using different parameterizations of the expansion around log-normal and fitting different combinations of energies, the discrepancy from log-normal (given by the value of the cumulant expansion at third order) for the energies from \citep{vonStecher:2009qw} varies between $\sim 17\% - 30\%$, which is approximately the same as the largest discrepancy between the energies from \citep{vonStecher:2009qw} and those predicted by an exactly log-normal distribution. Using similar expansions around log-normal and allowing for the $\sim 2\%$ deviation seen in the lattice data results in energies that differ from the log-normal predictions by no more than a few percent. 

I would like to emphasize that the log-normal distribution itself implies that these $2N$-body energies are universal, because all moments depend on a single energy scale, and are therefore not particularly sensitive to any non-universal effects. Thus, the lattice data, which shows a log-normal distribution, gives further confidence to the claim that there are universal $N$-body states tied to Efimov trimers. The log-normal distribution implies other universal properties as well. For example, if the two-body correlation function is log-normal, then by extension the correlation functions for all $2N$-body Efimov states must also be log-normal, but with rescaled parameters $\mu$ and $\sigma^2$. The logarithm of the $n$th moment of a $2N$-body correlator corresponds to $E_0^{(2Nn)} T = \frac{1}{2}N n (N n-1) E_0^{(4)} T$. Comparing this to the moments of the log-normal distribution (\Eq{lngenmoments}), gives $\mu = \frac{1}{2} N E_0^{(4)} T \, , \sigma^2 = -N^2 E_0^{(4)} T$ for the distribution of the $2N$-body correlation function. Thus, the distribution itself may be seen as universal for all correlation functions of states tied to Efimov trimers.

It is compelling that as the beta function for the two-particle coupling approaches an RG fixed point, the distribution for the two-particle correlator appears to as well. 
It may be interesting to explore an effective field theory type (or possibly some perturbative) expansion around the log-normal distribution \citep{Endres:2011mm}, much in the way that the KSW expansion \citep{Kaplan:1998tg,Kaplan:1996xu} for nuclear effective field theory expands around the unitary fixed point, either analytically using \Eq{char} or numerically using lattice data. Such a program, in addition to giving further insight into the connection between the log-normal distribution and unitarity, would allow calculations of non-universal effects at large $N$, as well as any possible deviations from log-normal within the universal regime.

\begin{acknowledgement}
The author would like to thank M. Endres, D. B. Kaplan, J.-W. Lee, J. E. Drut, P. Bedaque, T. Cohen, E. Berkowitz, K. McElvain, and T. Kurth for helpful discussions on this work.
This work was supported by the U.S. Department of Energy under grants DE-FG02-93ER-40762, DE-SC00046548.
\end{acknowledgement}

\bibliography{unitary}

\end{document}